\documentclass[journal, onecolumn, 12pt]{IEEEtran}

\usepackage{graphics}
\usepackage{rotating}
\usepackage{amsfonts}
\usepackage{amsmath}
\usepackage{subfigure}
\usepackage{epsfig}
\usepackage{enumerate}
\usepackage{setspace}
\usepackage{verbatim}

\graphicspath{{figures/}} \DeclareGraphicsExtensions{.eps,.ps}

\doublespace \centerfigcaptionstrue

\begin{document}

\title{Performance Analysis of the IEEE 802.11e Enhanced Distributed Coordination Function using Cycle Time Approach \footnotemark{$^{\dag}$}}

\author{\singlespace \normalsize \authorblockN{Inanc Inan, Feyza Keceli, and Ender Ayanoglu}\\
\authorblockA{Center for Pervasive Communications and Computing \\
Department of Electrical Engineering and Computer Science\\
The Henry Samueli School of Engineering\\
University of California, Irvine, 92697-2625\\
Email: \{iinan, fkeceli, ayanoglu\}@uci.edu}}

\maketitle

\footnotetext{$^{\dag}$ This work is supported by the Center for
Pervasive Communications and Computing, and by Natural Science
Foundation under Grant No. 0434928. Any opinions, findings, and
conclusions or recommendations expressed in this material are
those of authors and do not necessarily reflect the view of the
Natural Science Foundation.}

\begin{abstract}
The recently ratified IEEE 802.11e standard defines the Enhanced
Distributed Channel Access (EDCA) function for Quality-of-Service
(QoS) provisioning in the Wireless Local Area Networks (WLANs).
The EDCA uses Carrier Sense Multiple Access with Collision
Avoidance (CSMA/CA) and slotted Binary Exponential Backoff (BEB)
mechanism. We present a simple mathematical analysis framework for
the EDCA function. Our analysis considers the fact that the
distributed random access systems exhibit cyclic behavior where
each station successfully transmits a packet in a cycle. Our
analysis shows that an AC-specific cycle time exists for the EDCA
function. Validating the theoretical results via simulations, we
show that the proposed analysis accurately captures EDCA
saturation performance in terms of average throughput, medium
access delay, and packet loss ratio. The cycle time analysis is a
simple and insightful substitute for previously proposed more
complex EDCA models.
\end{abstract}

\section{Introduction}

The IEEE 802.11e standard \cite{802.11e} specifies the Hybrid
Coordination Function (HCF) which enables prioritized and
parameterized Quality-of-Service (QoS) services at the MAC layer.
The HCF combines a distributed contention-based channel access
mechanism, referred to as Enhanced Distributed Channel Access
(EDCA), and a centralized polling-based channel access mechanism,
referred to as HCF Controlled Channel Access (HCCA). We confine
our analysis to the EDCA scheme, which uses Carrier Sense Multiple
Access with Collision Avoidance (CSMA/CA) and slotted Binary
Exponential Backoff (BEB) mechanism as the basic access method.
The EDCA defines multiple Access Categories (AC) with AC-specific
Contention Window (CW) sizes, Arbitration Interframe Space (AIFS)
values, and Transmit Opportunity (TXOP) limits to support
MAC-level QoS and prioritization.

We evaluate the EDCA performance for the saturation (asymptotic)
case. The saturation analysis provides the limits reached by the
system throughput and protocol service time in stable conditions
when every station has always backlogged data ready to transmit in
its buffer. The analysis of the saturation provides in-depth
understanding and insights into the random access schemes and the
effects of different contention parameters on the performance. The
results of such analysis can be employed in access parameter
adaptation or in a call admission control algorithm.

Our analysis is based on the fact that a random access system
exhibits cyclic behavior. A cycle time is defined as the duration
in which an arbitrary tagged user successfully transmits one
packet on average \cite{Medepalli05}. We will derive the explicit
mathematical expression of the AC-specific EDCA cycle time. The
derivation considers the AIFS and CW differentiation by employing
a simple average collision probability analysis. We will use the
EDCA cycle time to predict the first moments of the saturation
throughput, the service time, and the packet loss probability. We
will show that the results obtained using the cycle time model
closely follow the accurate predictions of the previously proposed
more complex analytical models and simulation results. Our cycle
time analysis can serve as a simple and practical alternative
model for EDCA saturation throughput analysis.

\section{EDCA Overview}\label{sec:EDCAoverview}

The IEEE 802.11e EDCA is a QoS extension of IEEE 802.11
Distributed Coordination Function (DCF). The major enhancement to
support QoS is that EDCA differentiates packets using different
priorities and maps them to specific ACs that are buffered in
separate queues at a station. Each AC$_{i}$ within a station
($0\leq i\leq i_{max}$, $i_{max}=3$ in \cite{802.11e}) having its
own EDCA parameters contends for the channel independently of the
others. Following the convention of \cite{802.11e}, the larger the
index $i$ is, the higher the priority of the AC is. Levels of
services are provided through different assignments of the
AC-specific EDCA parameters; AIFS, CW, and TXOP limits.

If there is a packet ready for transmission in the MAC queue of an
AC, the EDCA function must sense the channel to be idle for a
complete AIFS before it can start the transmission. The AIFS of an
AC is determined by using the MAC Information Base (MIB)
parameters as $AIFS = SIFS + AIFSN \times T_{slot}$, where $AIFSN$
is the AC-specific AIFS number, $SIFS$ is the length of the Short
Interframe Space, and $T_{slot}$ is the duration of a time slot.

If the channel is idle when the first packet arrives at the AC
queue, the packet can be directly transmitted as soon as the
channel is sensed to be idle for AIFS. Otherwise, a backoff
procedure is completed following the completion of AIFS before the
transmission of this packet. A uniformly distributed random
integer, namely a backoff value, is selected from the range
$[0,W]$.
The backoff counter is decremented at the slot boundary if the
previous time slot is idle. Should the channel be sensed busy at
any time slot during AIFS or backoff, the backoff procedure is
suspended at the current backoff value. The backoff resumes as
soon as the channel is sensed to be idle for AIFS again. When the
backoff counter reaches zero, the packet is transmitted in the
following slot.

The value of $W$ depends on the number of retransmissions the
current packet experienced. The initial value of $W$ is set to the
AC-specific $CW_{min}$. If the transmitter cannot receive an
Acknowledgment (ACK) packet from the receiver in a timeout
interval, the transmission is labeled as unsuccessful and the
packet is scheduled for retransmission. At each unsuccessful
transmission, the value of $W$ is doubled until the maximum
AC-specific $CW_{max}$ limit is reached. The value of $W$ is reset
to the AC-specific $CW_{min}$ if the transmission is successful,
or the retry limit is reached thus the packet is dropped.

The higher priority ACs are assigned smaller AIFSN. Therefore, the
higher priority ACs can either transmit or decrement their backoff
counters while lower priority ACs are still waiting in AIFS. This
results in higher priority ACs facing a lower average probability
of collision and relatively faster progress through backoff slots.
Moreover, in EDCA, the ACs with higher priority may select backoff
values from a comparably smaller CW range. This approach
prioritizes the access since a smaller CW value means a smaller
backoff delay before the transmission.

Upon gaining the access to the medium, each AC may carry out
multiple frame exchange sequences as long as the total access
duration does not go over a TXOP limit. Within a TXOP, the
transmissions are separated by SIFS. Multiple frame transmissions
in a TXOP can reduce the overhead due to contention. A TXOP limit
of zero corresponds to only one frame exchange per access.

An internal (virtual) collision within a station is handled by
granting the access to the AC with the highest priority. The ACs
with lower priority that suffer from a virtual collision run the
collision procedure as if an outside collision has occured.

\section{Related Work}

In this section, we provide a brief summary of the studies in the
literature on the theoretical DCF and EDCA function saturation
performance analysis.

Three major saturation performance models have been proposed for
DCF; \textit{i)} assuming constant collision probability for each
station, Bianchi \cite{Bianchi00} developed a simple Discrete-Time
Markov Chain (DTMC) and the saturation throughput is obtained by
applying regenerative analysis to a generic slot time,
\textit{ii)} Cali \textit{et al.} \cite{Cali00} employed renewal
theory to analyze a \textit{p}-persistent variant of DCF with
persistence factor \textit{p} derived from the CW, and
\textit{iii)} Tay \textit{et al.} \cite{Tay01} instead used an
average value mathematical method to model DCF backoff procedure
and to calculate the average number of interruptions that the
backoff timer experiences. Having the common assumption of slot
homogeneity (for an arbitrary station, constant collision or
transmission probability at an arbitrary slot), these models
define all different renewal cycles all of which lead to accurate
saturation performance analysis.

These major methods (especially \cite{Bianchi00}) are modified by
several researchers to include the extra features of the EDCA
function in the saturation analysis. Xiao \cite{Xiao05} extended
\cite{Bianchi00} to analyze only the CW differentiation. Kong
\textit{et al.} \cite{Kong04} took AIFS differentiation into
account. On the other hand, these EDCA extensions miss the
treatment of varying collision probabilities at different AIFS
slots due to varying number of contending stations. Robinson
\textit{et al.} \cite{Robinson04} proposed an average analysis on
the collision probability for different contention zones during
AIFS. Hui \textit{et al.} \cite{Hui05} unified several major
approaches into one approximate average model taking into account
varying collision probability in different backoff subperiods
(corresponds to contention zones in \cite{Robinson04}). Zhu
\textit{et al.} \cite{Zhu05} proposed another analytical EDCA
Markov model averaging the transition probabilities based on the
number and the parameters of high priority flows. Inan \textit{et
al.} \cite{Inan07_ICC} proposed a 3-dimensional DTMC which
provides accurate treatment of AIFS and CW differentiation.
Another 3-dimensional DTMC is proposed by Tao \textit{et al.}
\cite{Tao06} in which the third dimension models the state of
backoff slots between successive transmission periods. The fact
that the number of idle slots between successive transmissions can
be at most the minimum of AC-specific $CW_{max}$ values is
considered. Independently, Zhao \textit{et al.} \cite{Zhao02} had
previously proposed a similar model for the heterogeneous case
where each station has traffic of only one AC. Banchs \textit{et
al.} \cite{Banchs06} proposed another model which considers
varying collision probability among different AIFS slots due to a
variable number of stations. Lin \textit{et al.} \cite{Lin06}
extended \cite{Tay01} in order to carry out mean value analysis
for approximating AIFS and CW differentiation.

Our approach is based on the observation that the transmission
behavior in the 802.11 WLAN follows a pattern of periodic cycles.
Previously, Medepalli \textit{et al.} \cite{Medepalli05} provided
explicit expressions for average DCF cycle time and system
throughput. Similarly, Kuo \textit{et al.} \cite{Kuo03} calculated
the EDCA transmission cycle assuming constant collision
probability for any traffic class. On the other hand, such an
assumption leads to analytical inaccuracies
\cite{Kong04}-\cite{Lin06}. The main contribution is that we
incorporate accurate AIFS and CW differentiation calculation in
the EDCA cycle time analysis. We show that the cyclic behavior is
observed on a per AC basis in the EDCA. To maintain the simplicity
of the cycle time analysis, we employ averaging on the AC-specific
collision probability. The comparison with more complex and
detailed theoretical and simulation models reveals that the
analytical accuracy is preserved.

\section{EDCA Cycle Time Analysis}

In this section, we will first derive the AC-specific average
collision probability. Next, we will calculate the AC-specific
average cycle time. Finally, we will relate the average cycle time
and the average collision probability to the average normalized
throughput, EDCA service time, and packet loss probability.

\subsection{AC-specific Average Collision Probability}

\begin{figure}
\center{\epsfig{file=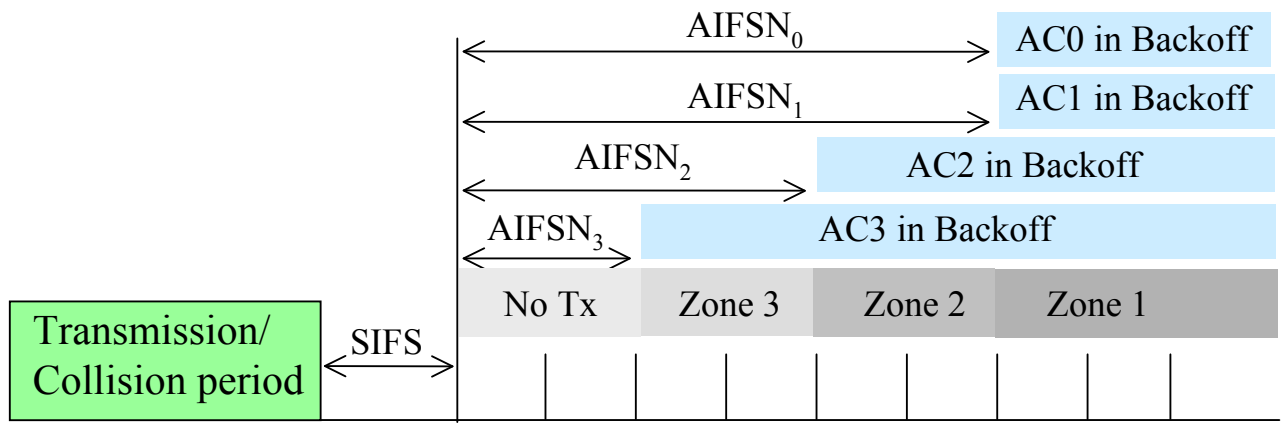,height=17
cm,angle=-90}} \caption[r] {\label{fig:unsat_contzones} EDCA
backoff after busy medium.}
\end{figure}

The difference in AIFS of each AC in EDCA creates the so-called
\textit{contention zones or periods} as shown in
Fig.~\ref{fig:unsat_contzones} \cite{Robinson04},\cite{Hui05}. In
each contention zone, the number of contending stations may vary.
We employ an average analysis on the AC-specific collision
probability rather than calculating it separately for different
AIFS and backoff slots as in \cite{Inan07_ICC}-\cite{Banchs06}. We
calculate the AC-specific collision probability according to the
long term occupancy of AIFS and backoff slots.

We define $p_{c_{i,x}}$ as the conditional probability that
AC$_{i}$ experiences either an external or an internal collision
given that it has observed the medium idle for $AIFS_{x}$ and
transmits in the current slot (note $AIFS_{x}\geq AIFS_{i}$ should
hold). For the following, in order to be consistent with the
notation of \cite{802.11e}, we assume $AIFS_{0}\geq AIFS_{1} \geq
AIFS_{2} \geq AIFS_{3}$. Let $d_{i} = AIFSN_{i} - AIFSN_{3}$.
Following the slot homogeneity assumption of \cite{Bianchi00},
assume that each AC$_{i}$ transmits with constant probability,
$\tau_{i}$. Also, let the total number AC$_{i}$ flows be $N_{i}$.
Then, for the heterogeneous scenario in which each station has
only one AC
\begin{equation}
\label{eq:unsatpcix} \setlength{\nulldelimiterspace}{0pt}
p_{c_{i,x}} = 1-\frac{\prod \limits_{i':d_{i'}\leq d_{x}}
(1-\tau_{i'})^{N_{i'}}}{(1-\tau_{i})}.
\end{equation}
\noindent We only formulate the situation when there is only one
AC per station, therefore no internal collisions can occur. Note
that this simplification does not cause any loss of generality,
because the proposed model can be extended for the case of higher
number of ACs per station as in \cite{Kong04},\cite{Inan07_ICC}.

\begin{figure}
\center{\epsfig{file=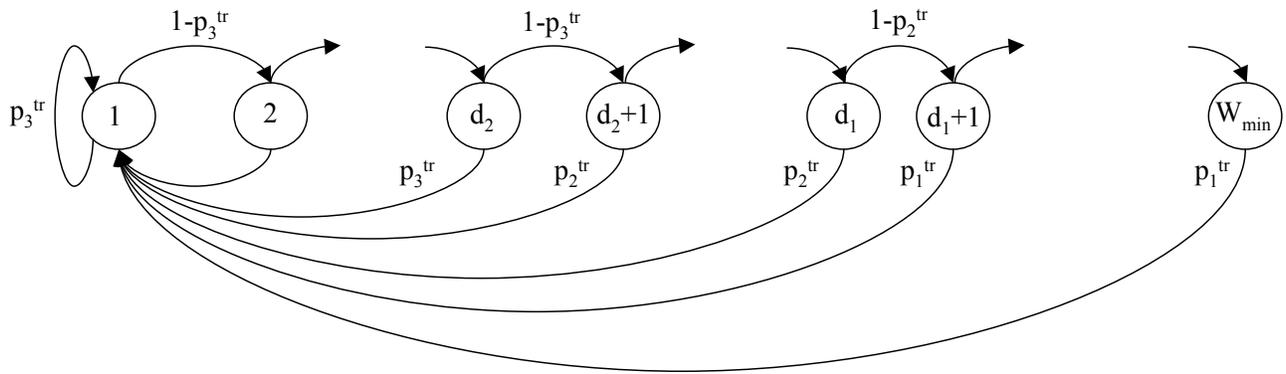,height=17
cm,angle=-90}} \caption[r] {\label{fig:unsat_AIFSMC} Transition
through backoff slots in different contention zones for the
example given in Fig.\ref{fig:unsat_contzones}.}
\end{figure}

We use the Markov chain shown in Fig.~\ref{fig:unsat_AIFSMC} to
find the long term occupancy of the contention zones. Each state
represents the $n^{th}$ backoff slot after the completion of the
AIFS$_{3}$ idle interval following a transmission period. The
Markov chain model uses the fact that a backoff slot is reached if
and only if no transmission occurs in the previous slot. Moreover,
the number of states is limited by the maximum idle time between
two successive transmissions which is $W_{min}=\min(CW_{i,max})$
for a saturated scenario. The probability that at least one
transmission occurs in a backoff slot in contention zone $x$ is
\begin{equation}
\label{eq:unsatptr} \setlength{\nulldelimiterspace}{0pt}
p^{tr}_{x} = 1-\prod_{i':d_{i'}\leq d_{x}} (1-\tau_{i'})^{N_{i'}}.
\end{equation}
\noindent Note that the contention zones are labeled with $x$
regarding the indices of $d$. In the case of an equality in AIFS
values of different ACs, the contention zone is labeled with the
index of AC with higher priority.

Given the state transition probabilities as in
Fig.~\ref{fig:unsat_AIFSMC}, the long term occupancy of the
backoff slots $b'_{n}$ can be obtained from the steady-state
solution of the Markov chain. Then, the AC-specific average
collision probability $p_{c_{i}}$ is found by weighing zone
specific collision probabilities $p_{c_{i,x}}$ according to the
long term occupancy of contention zones (thus backoff slots)
\begin{equation}
\label{eq:unsatpci}p_{c_{i}} = \frac{\sum_{n=d_{i}+1}^{W_{min}}
p_{c_{i,x}}b'_{n}}{\sum_{n=d_{i}+1}^{W_{min}} b'_{n}}
\end{equation}
\noindent where $x = \max \left( y~|~d_{y} = \underset{z}{\max}
(d_{z}~|~d_{z} \leq n)\right)$ which shows $x$ is assigned the
highest index value within a set of ACs that have AIFSN smaller
than or equal to $n+AIFSN_{3}$. This ensures that at backoff slot
$n$, AC$_{i}$ has observed the medium idle for AIFS$_{x}$.
Therefore, the calculation in~(\ref{eq:unsatpci}) fits into the
definition of $p_{c_{i,x}}$.

\subsection{AC-Specific Average Cycle Time}

Intuitively, it can be seen that each user transmitting at the
same AC has equal cycle time, while the cycle time may differ
among ACs. Our analysis will also mathematically show this is the
case. Let $E_{i}[t_{cyc}]$ be average cycle time for a tagged
AC$_{i}$ user. $E_{i}[t_{cyc}]$ can be calculated as the sum of
average duration for \textit{i)} the successful transmissions,
$E_{i}[t_{suc}]$, \textit{ii)} the collisions, $E_{i}[t_{col}]$,
and \textit{iii)} the idle slots, $E_{i}[t_{idle}]$ in one cycle.

In order to calculate the average time spent on successful
transmissions during an AC$_{i}$ cycle time, we should find the
expected number of total successful transmissions between two
successful transmissions of AC$_{i}$. Let $Q_{i}$ represent this
random variable. Also, let $\gamma_{i}$ be the probability that
the transmitted packet belongs to an arbitrary user from AC$_{i}$
given that the transmission is successful. Then,
\begin{equation}\label{eq:gamma_i}
\gamma_{i} = \sum_{n=d_{i}+1}^{W_{min}}
b'_{n}\frac{p_{s_{i,n}}/N_{i}}{\sum \limits_{\forall j}
p_{s_{j,n}}}
\end{equation}
\noindent where
\begin{equation}\label{eq:p_s_i_cycle}
p_{s_{i,n}} =
\left\{ \\
\begin{IEEEeqnarraybox}[\relax][c]{lc}
\frac{N_{i}\tau_{i}}{(1-\tau_{i})}\prod_{i':d_{i'}\leq
n-1}(1-\tau{i'})^{N_{i'}}, &~{\rm if}~n \geq d_{i}+1 \\ 0, &~{\rm
if }~n < d_{i}+1.
\end{IEEEeqnarraybox}
\right.
\end{equation}


Then, the Probability Mass Function (PMF) of $Q_{i}$ is
\begin{equation}\label{eq:PMFsucctrans}
Pr(Q_{i}=k) = \gamma_{i}(1-\gamma_{i})^{k}, ~~k \geq 0.
\end{equation}

We can calculate expected number of successful transmissions of
any AC$_{j}$ during the cycle time of AC$_{i}$, $ST_{j,i}$, as
\begin{equation}\label{eq:ExpectedindividualAC}
ST_{j,i} = N_{j}E[Q_{i}] \frac{\gamma_{j}}{1-\gamma_{i}}.
\end{equation}

Inserting $E[Q_{i}]=(1-\gamma_{i})/\gamma_{i}$ in
(\ref{eq:ExpectedindividualAC}), our intuition that each user from
AC$_{i}$ can transmit successfully once on average during the
cycle time of another AC$_{i}$ user, i.e., $ST_{i,i}=N_{i}$, is
confirmed. Therefore, the average cycle time of any user belonging
to the same AC is equal in a heterogeneous scenario where each
station runs only one AC. Including the own successful packet
transmission time of tagged AC$_{i}$ user in $E_{i}[t_{suc}]$, we
find
\begin{equation}\label{eq:Etsuc}
E_{i}[t_{suc}] = \sum_{\forall j} ST_{j,i}T_{s_{j}}
\end{equation}

\noindent where $T_{s_{j}}$ is defined as the time required for a
successful packet exchange sequence. $T_{s_{j}}$ will be derived
in (\ref{eq:unsatTs}).

To obtain $E_{i}[t_{col}]$, we need to calculate average number of
users that involve in a collision, $N_{c_{n}}$, at the $n^{th}$
slot after last busy time for given $N_{i}$ and $\tau_{i}$,
$\forall i$. Let the total number of users transmitting at the
$n^{th}$ slot after last busy time be denoted as $Y_{n}$. We see
that $Y_{n}$ is the sum of random variables,
$Binomial(N_{i},\tau_{i})$, $\forall i:~d_{i}\leq n-1$. Employing
simple probability theory, we can calculate
$N_{c_{n}}=E[Y_{n}|Y_{n}\geq 2]$. After some simplification,
\begin{equation}
N_{c_{n}} = \frac{\sum\limits_{i:d_{i}\leq n-1}
(N_{i}\tau_{i}-p_{s_{i,n}})}{1-\prod\limits_{i:d_{i}\leq
n-1}(1-\tau_{i})^{N_{i}}-\sum\limits_{i:d_{i}\leq n-1}p_{s_{i,n}}}
\end{equation}

If we let the average number of users involved in a collision at
an arbitrary backoff slot be $N_{c}$, then
\begin{equation}
N_{c} = \sum_{\forall n} b'_{n}N_{c_{n}}.
\end{equation}

We can also calculate the expected number of collisions that an
AC$_{j}$ user experiences during the cycle time of an AC$_{i}$,
$CT_{j,i}$, as
\begin{equation}
CT_{j,i} = \frac{p_{c_{j}}}{1-p_{c_{j}}}ST_{j,i}.
\end{equation}

\noindent Then, defining $T_{c_{j}}$ as the time wasted in a
collision period (will be derived in (\ref{eq:unsatTc}),
\begin{equation}
E_{i}[t_{col}] = \frac{1}{N_{c}}\sum_{\forall j}
CT_{j,i}T_{c_{j}}.
\end{equation}

Given $p_{c_{i}}$, we can calculate the expected number of backoff
slots $E_{i}[t_{bo}]$ that AC$_{i}$ waits before attempting a
transmission. Let $W_{i,k}$ be the CW size of AC$_{i}$ at backoff
stage $k$ \cite{Inan07_ICC}. Note that, when the retry limit
$r_{i}$ is reached, any packet is discarded. Therefore, another
$E_{i}[t_{bo}]$ passes between two transmissions with probability
$p_{c_{i}}^{r_{i}}$
\begin{equation}\label{eq:aveBO}
E_{i}[t_{bo}]=\frac{1}{1-p_{c_{i}}^{r_{i}}}\sum_{k=1}^{r}p_{c_{i}}^{k-1}(1-p_{c_{i}})\frac{W_{i,k}}{2}.
\end{equation}

\noindent Noticing that between two successful transmissions,
AC$_{i}$ also experiences $CT_{i,i}$ collisions,
\begin{equation}\label{eq:E_i_t_idle}
E_{i}[t_{idle}] = E_{i}[t_{bo}](CT_{i,i}/N_{i}+1)t_{slot}.
\end{equation}

As shown in \cite{Hui05}, the transmission probability of a user
using AC$_{i}$,
\begin{equation}\label{eq:tauapp}
\tau_{i} = \frac{1}{E_{i}[t_{bo}]+1}.
\end{equation}

Note that, in \cite{Hui05}, it is proven that the mean value
analysis for the average transmission probability as in
(\ref{eq:tauapp}) matches the Markov analysis of \cite{Bianchi00}.

The fixed-point equations (\ref{eq:unsatpcix})-(\ref{eq:tauapp})
can numerically be solved for $\tau_{i}$ and $p_{c_{i}}$, $\forall
i$. Then, each component of the average cycle time for AC$_{i}$,
$\forall i$, can be calculated using
(\ref{eq:gamma_i})-(\ref{eq:E_i_t_idle}).

\subsection{Performance Analysis}

Let $T_{p_{i}}$ be the average payload transmission time for
AC$_{i}$ ($T_{p_{i}}$ includes the transmission time of MAC and
PHY headers), $\delta$ be the propagation delay, $T_{ack}$ be the
time required for acknowledgment packet (ACK) transmission. Then,
for the basic access scheme, we define the time spent in a
successful transmission $T_{s_{i}}$ and a collision $T_{c_{i}}$
for any AC$_{i}$ as
\begin{align}\label{eq:unsatTs}
T_{s_{i}} = & T_{p_{i}} + \delta + SIFS + T_{ack} + \delta +
AIFS_{i}
\\ \label{eq:unsatTc} T_{c_{i}} = & T_{p^{*}_{i}} + ACK\_Timeout +
AIFS_{i}
\end{align}
\noindent where $T_{p^{*}_{i}}$ is the average transmission time
of the longest packet payload involved in a collision
\cite{Bianchi00}. For simplicity, we assume the packet size to be
equal for any AC, then $T_{p^{*}_{i}}=T_{p_{i}}$. Being not
explicitly specified in the standards, we set $ACK\_Timeout$,
using Extended Inter Frame Space (EIFS) as $EIFS_{i}-AIFS_{i}$.
Note that the extensions of~(\ref{eq:unsatTs})
and~(\ref{eq:unsatTc}) for the RTS/CTS scheme are straightforward
\cite{Bianchi00}.

The average cycle time of an AC represents the renewal cycle for
each AC. Then, the normalized throughput of AC$_{i}$ is defined as
the successfully transmitted information per renewal cycle
\begin{equation}\label{eq:Si_cycle}
S_{i} =
\frac{N_{i}T_{p_{i}}}{E_{i}[t_{suc}]+E_{i}[t_{col}]+E_{i}[t_{idle}]}.
\end{equation}

The AC-specific cycle time is directly related but not equal to
the mean protocol service time. By definition, the cycle time is
the duration between successful transmissions. We define the
average protocol service time such that it also considers the
service time of packets which are dropped due to retry limit. On
the average, $1/p_{i,drop}$ service intervals correspond to
$1/p_{i,drop}-1$ cycles. Therefore, the mean service time
$\mu_{i}$ can be calculated as
\begin{align}\label{eq:pdp_cycle}
\mu_{i} = (1-p_{i,drop})E_{i}[t_{cyc}].
\end{align}

Simply, the average packet drop probability due to MAC layer
collisions is
\begin{align}\label{eq:pidrop_cycle}
p_{i,drop} = p_{c_{i}}^{r_{i}}.
\end{align}

\section{Numerical and Simulation Results} \label{sec:Validation}

We validate the accuracy of the numerical results by comparing
them to the simulation results obtained from ns-2 \cite{ns2}. For
the simulations, we employ the IEEE 802.11e HCF MAC simulation
model for ns-2.28 \cite{ourcode}. This module implements all the
EDCA and HCCA functionalities stated in \cite{802.11e}.

In simulations, we consider two ACs, one high priority (AC$_{3}$)
and one low priority (AC$_{1}$). Each station runs only one AC.
Each AC has always buffered packets that are ready for
transmission. For both ACs, the payload size is 1000 bytes.
RTS/CTS handshake is turned on. The simulation results are
reported for the wireless channel which is assumed to be not prone
to any errors during transmission. The errored channel case is
left for future study. All the stations have 802.11g Physical
Layer (PHY) using 54 Mbps and 6 Mbps as the data and basic rate
respectively ($T_{slot}=9~\mu s$, $SIFS=10~\mu s$) \cite{802.11g}.
The simulation runtime is 100 seconds.

\begin{figure}
\centering \includegraphics[width = 1.0\linewidth]{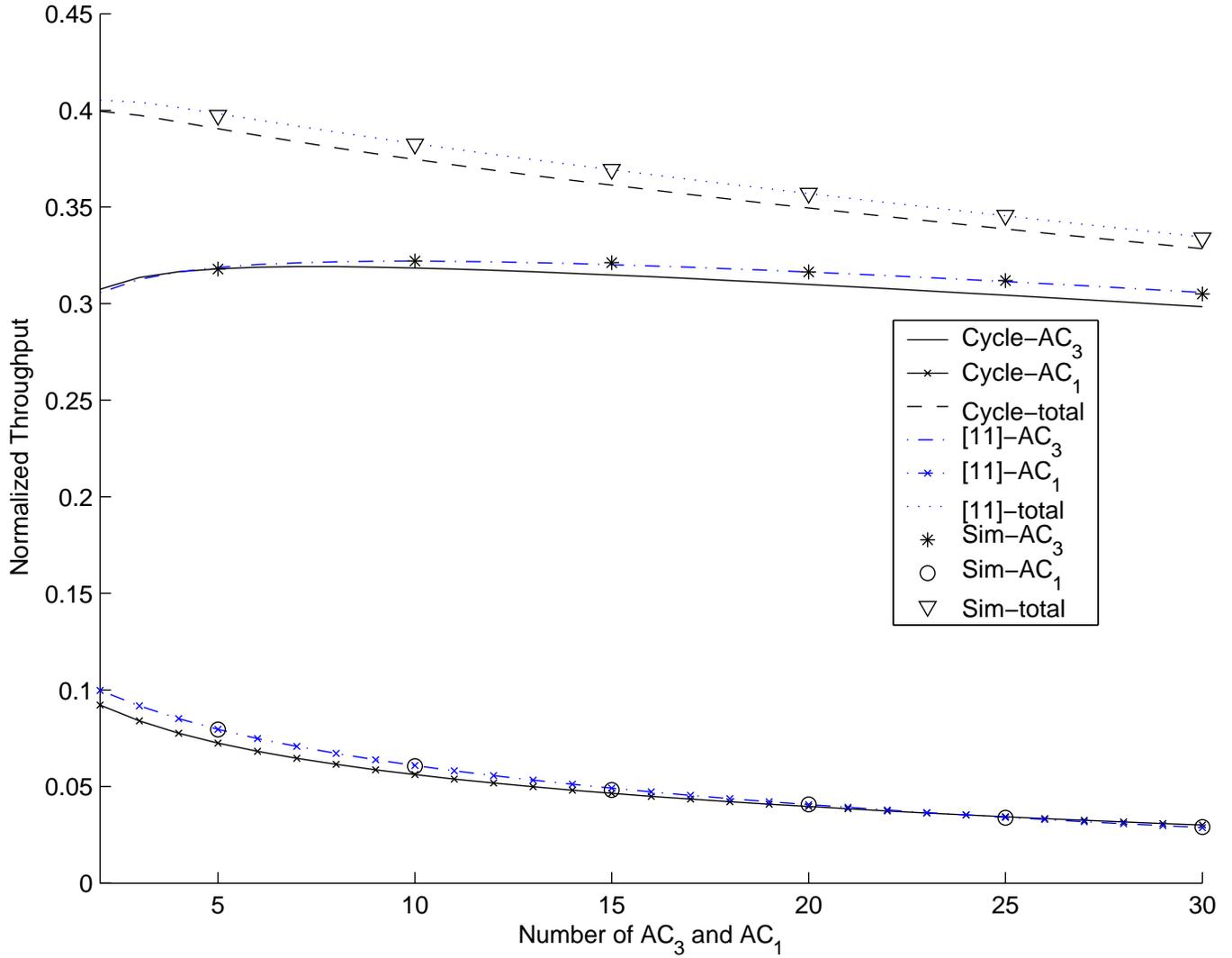}
\caption{Analyzed and simulated normalized throughput of each AC
when both $N_{1}$ and $N_{3}$ are varied from 5 to 30 and equal to
each other for the cycle time analysis. Analytical results of the
model proposed in \cite{Inan07_ICC} are also added for
comparison.} \label{fig:A1_thp_GC07}
\end{figure}

\begin{figure}[t]
\centering \includegraphics[width = 1.0\linewidth]{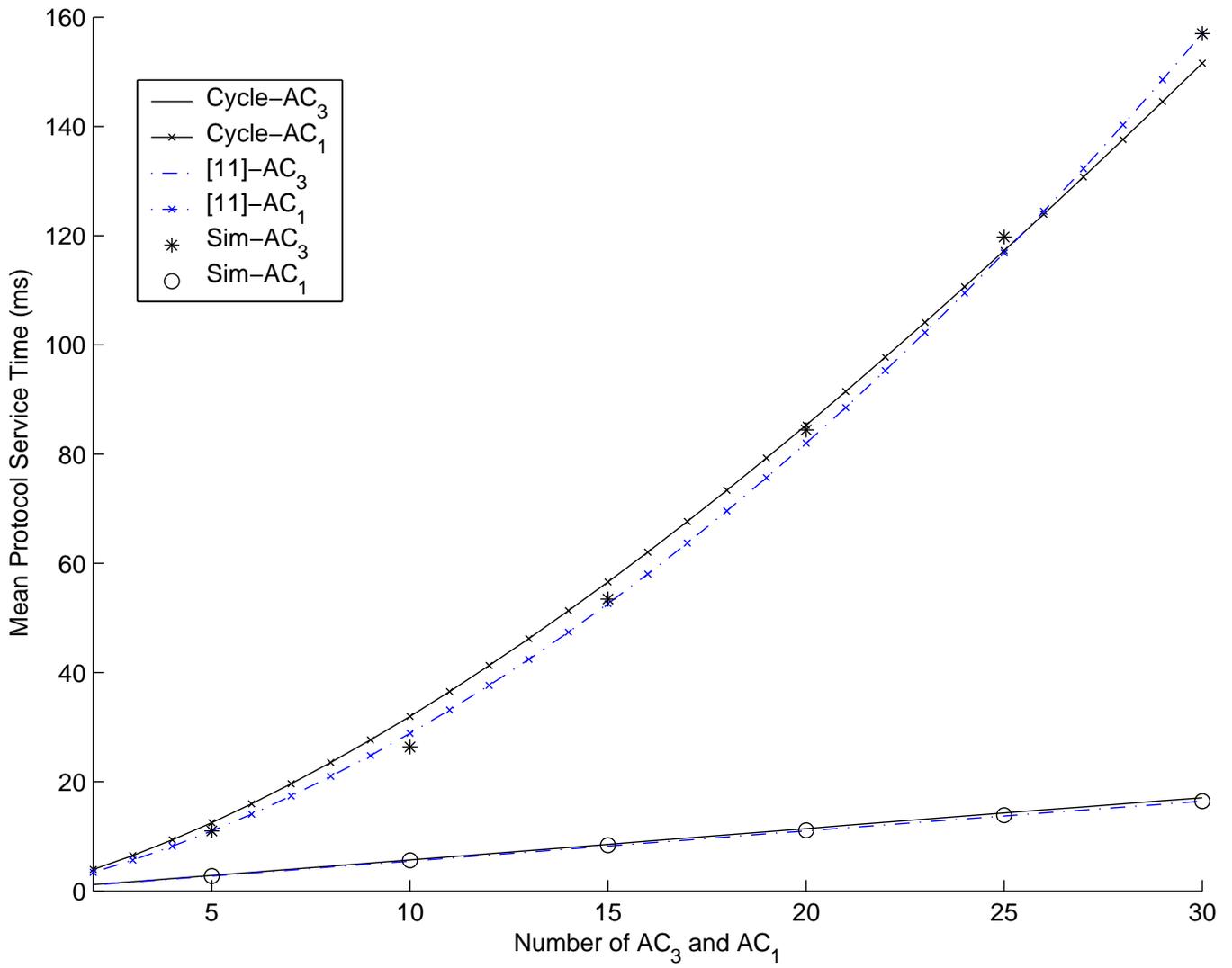}
\caption{Analyzed and simulated mean protocol service time of each
AC when both $N_{1}$ and $N_{3}$ are varied from 5 to 30 and equal
to each other for the proposed cycle time analysis and the model
in \cite{Inan07_ICC}.} \label{fig:A1_mpst_GC07}
\end{figure}

\begin{figure}[t]
\centering \includegraphics[width = 1.0\linewidth]{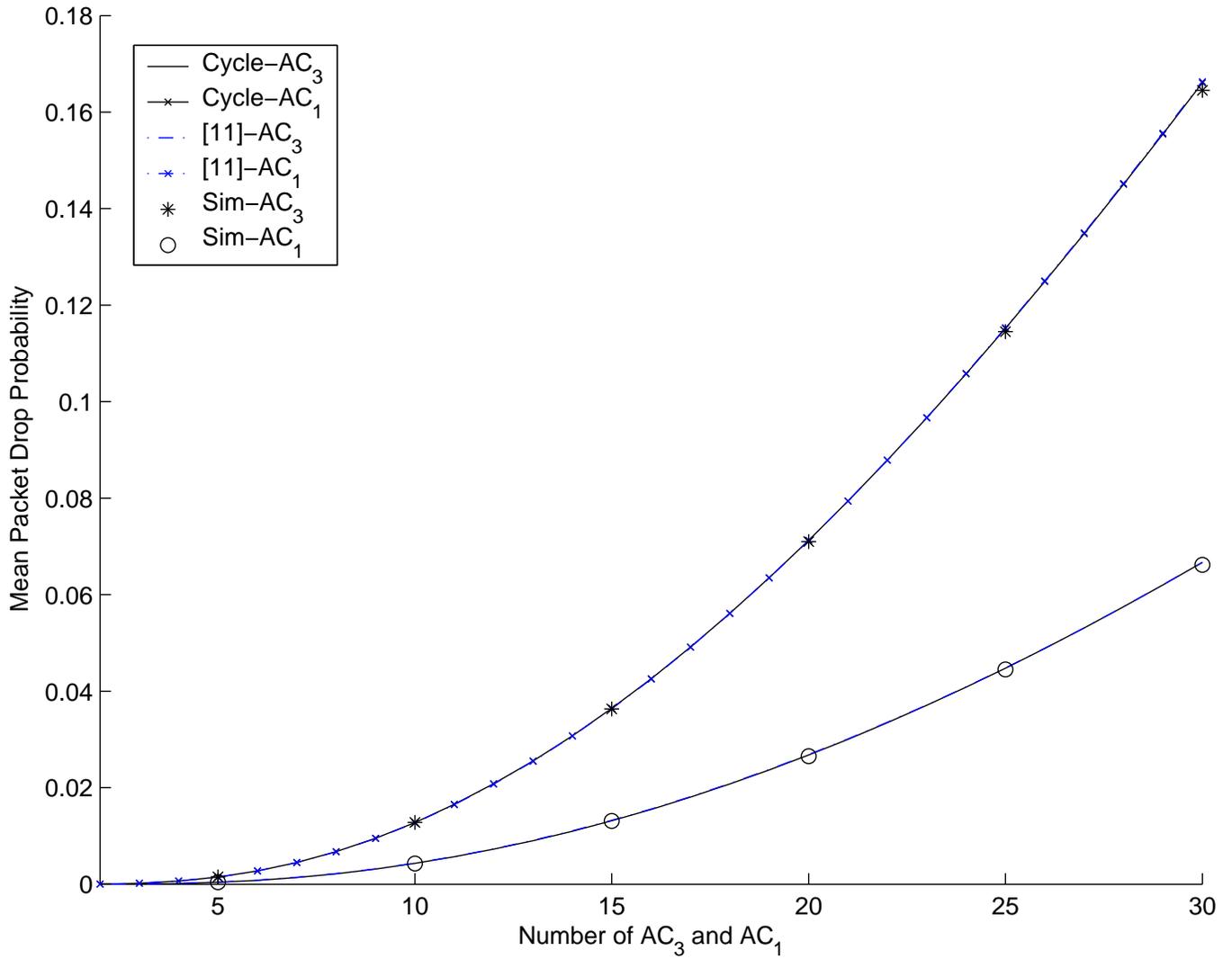}
\caption{Analyzed and simulated mean packet drop probability of
each AC when both $N_{1}$ and $N_{3}$ are varied from 5 to 30 and
equal to each other for the proposed cycle time analysis and the
model in \cite{Inan07_ICC}.} \label{fig:A1_mpdp_GC07}
\end{figure}

In the first set of experiments, we set $AIFSN_{1}=3$,
$AIFSN_{3}=2$, $CW_{1,min}=31$, $CW_{3,min}=15$, $m_{1}=m_{3}=3$,
$r_{1}=r_{3}=7$. Fig.~\ref{fig:A1_thp_GC07} shows the normalized
throughput of each AC when both $N_{1}$ and $N_{3}$ are varied
from 5 to 30 and equal to each other. As the comparison with a
more detailed analytical model \cite{Inan07_ICC} and the
simulation results reveal, the cycle time analysis can predict
saturation throughput accurately.
Fig.~\ref{fig:A1_mpst_GC07} and Fig.~\ref{fig:A1_mpdp_GC07}
display the mean protocol service time and packet drop probability
respectively for the same scenario of Fig.~\ref{fig:A1_thp_GC07}.
As comparison with \cite{Inan07_ICC} and the simulation results
show, both performance measures can accurately be predicted by the
proposed cycle time model. Although not included in the figures, a
similar discussion holds for the comparison with other detailed
and/or complex models of \cite{Tao06}-\cite{Banchs06}.

\begin{figure}[t]
\centering \includegraphics[width =
1.0\linewidth]{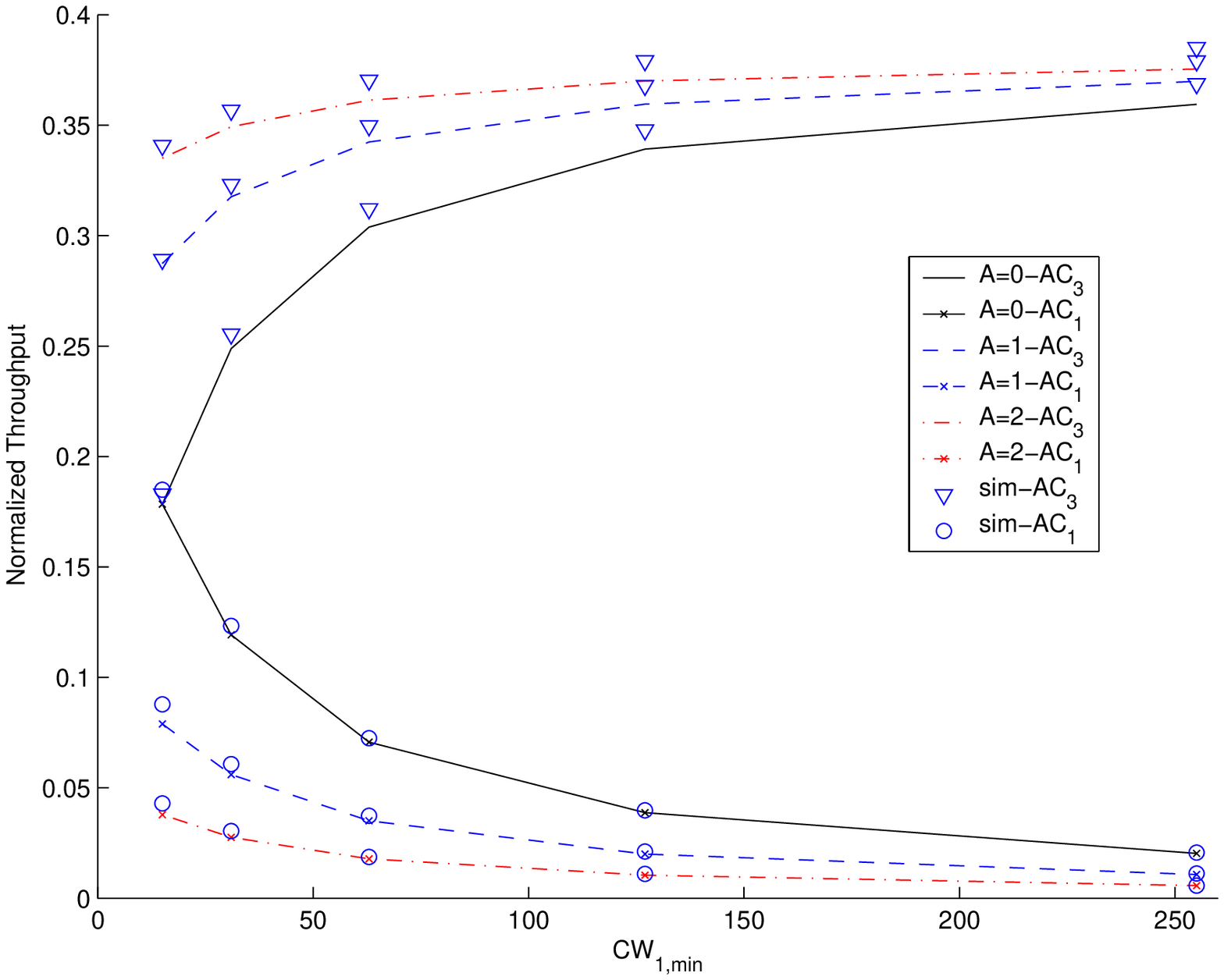} \caption{Analytically calculated
and simulated performance of each AC when $AIFSN_{3}=2$,
$CW_{3,min}=15$, $N_{1}=N_{3}=10$, $AIFSN_{1}$ varies from 2 to 4,
and $CW_{1,min}$ takes values from the set $\{15,31,63,127,255\}$.
Note that $AIFSN_{1}-AIFSN_{3}$ is denoted by $A$.}
\label{fig:v_aifs_cw_1_thp_GC07}
\end{figure}

\begin{figure}[t]
\centering \includegraphics[width =
1.0\linewidth]{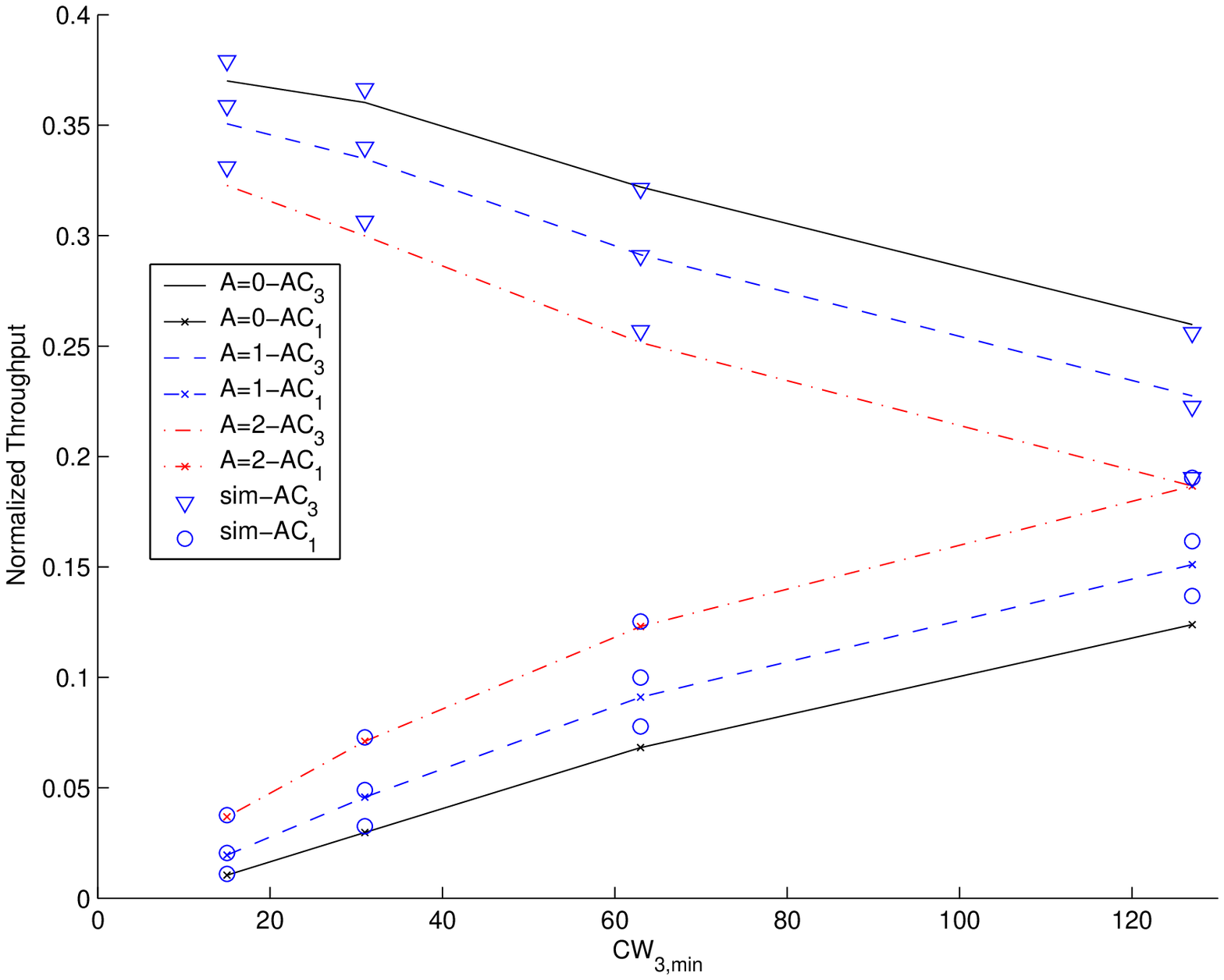} \caption{Analytically calculated
and simulated performance of each AC when $AIFSN_{1}=4$,
$CW_{1,min}=127$, $N_{1}=N_{3}=10$, $AIFSN_{3}$ varies from 2 to
4, and $CW_{3,min}$ takes values from the set $\{15,31,63,127\}$.
Note that $AIFSN_{1}-AIFSN_{3}$ is denoted by $A$.}
\label{fig:v_aifs_cw_3_thp_GC07}
\end{figure}

In the second set of experiments, we fix the EDCA parameters of
one AC and vary the parameters of the other AC in order to show
the proposed cycle time model accurately captures the normalized
throughput for different sets of EDCA parameters. In the
simulations, both $N_{1}$ and $N_{3}$ are set to 10.
Fig.~\ref{fig:v_aifs_cw_1_thp_GC07} shows the normalized
throughput of each AC when we set $AIFSN_{3}=2$, $CW_{3,min}=15$,
and vary $AIFSN_{1}$ and $CW_{1,min}$.
Fig.~\ref{fig:v_aifs_cw_3_thp_GC07} shows the normalized
throughput of each AC when we set $AIFSN_{1}=4$, $CW_{1,min}=127$,
and vary $AIFSN_{3}$ and $CW_{3,min}$. As the comparison with
simulation results show, the predictions of the proposed cycle
time model are accurate. We do not include the results for packet
drop probability and service time for this experiment. No
discernable trends toward error are observed.

\section{Conclusion}

We have presented an accurate cycle time model for predicting the
EDCA saturation performance analytically. The model accounts for
AIFS and CW differentiation mechanisms of EDCA. We employ a simple
average collision probability calculation regarding AIFS and CW
differentiation mechanisms of EDCA. Instead of generic slot time
analysis of \cite{Bianchi00}, we use the AC-specific cycle time as
the renewal cycle. We show that the proposed simple cycle time
model performs as accurate as more detailed and complex models
previously proposed in the literature. The mean saturation
throughput, protocol service time and packet drop probability are
calculated using the model. This analysis also highlights some
commonalities between approaches in EDCA saturation performance
analysis. The simple cycle time analysis can provide invaluable
insights for QoS provisioning in the WLAN.


\bibliographystyle{IEEEtran}
\bibliography{IEEEabrv,C:/INANCINAN/bibliography/standards,C:/INANCINAN/bibliography/HCCA,C:/INANCINAN/bibliography/simulations,C:/INANCINAN/bibliography/channel,C:/INANCINAN/bibliography/books,C:/INANCINAN/bibliography/mypapers,C:/INANCINAN/bibliography/myreports,C:/INANCINAN/bibliography/EDCAanalysis}

\clearpage

\clearpage

\clearpage

\clearpage

\clearpage

\clearpage

\clearpage

\end{document}